\documentclass[rnote]{aa} 
\usepackage{graphicx,natbib}
\bibpunct{(}{)}{;}{a}{}{,}
\usepackage{txfonts}

\def\kms{$\mbox{km s}^{-1}$}
\begin{document}
   \title{A way to deal with the fringe-like pattern in VIMOS-IFU data \thanks{Based on observations taken at the La Silla Paranal observatory withing Program ID: 079.B-0402}} 

   \subtitle{}

   \author{C. Lagerholm\inst{1} \and H. Kuntschner\inst{1}      \and M. Cappellari\inst{2}  \and D. Krajnovi\'c\inst{1} \and R. M. McDermid\inst{3}  \and M. Rejkuba\inst{1} }

   \institute{European Southern Observatory, Karl-Schwarzschild-Str 2, 85748 Garching bei M\"{u}nchen, Germany\\
              \email{clagerho@eso.org}
         \and
             Sub-department of Astrophysics, University of Oxford, Denys Wilkinson Building, Keble Road, Oxford OX1 3RH
         \and
             Gemini Observatory, Northern Operations Center, 670 N. A'ohoku Place, Hilo, HI 96720, USA}

   \date{Received 31 January 2012; Accepted 12 Mars 2012 }
 
  \abstract
  {The use of integral field units (IFUs) is now commonplace at all
    major observatories offering efficient means of obtaining spectral
    as well as imaging information at the same time. The IFU instrument
    designs are complex and spectral images have typically highly
    condensed formats, therefore present challenges for the IFU
    data reduction pipelines.  In the case of the VLT VIMOS-IFU, a
    fringe-like pattern affecting the spectra well into the optical
    and blue wavelength regime as well as artificial intensity
    variations, require additional reduction steps beyond standard
    pipeline processing. }
%
%
  {We propose an empirical method for the
    removal of the fringe-like pattern in the spectral domain and the
    intensity variations in the imaging domain. We also demonstrate
    the potential consequences of a failure to correct for these effects for any subsequent data analysis. Here we use the example of deriving stellar
    velocity, velocity dispersion, and absorption line-strength maps for
    early-type galaxies.}
%
%
  {For each spectrum that we reduce with the ESO standard VIMOS
    pipeline, we derive a correction spectrum by using the median of the eight
    surrounding spectra as a proxy for the unaffected, underlying
    spectrum. This method relies on the fact that our science targets
    (nearby early-type galaxies) cover the complete field-of-view of
    the VIMOS-IFU, have spectral properties that vary gradually with position and that the
    exact shape of the fringe-like pattern is nearly independent and highly
    variable between neighboring spatial positions.
    Quadrant-to-quadrant intensity variations are corrected for in an
    independent step. }
 %
  {We find that the proposed correction methods for the removal of the
    fringe-like pattern and the intensity variations in VIMOS-IFU
    data-cubes are suitable to permit a meaningful data analysis for
    our sample of nearby early-type galaxies. Since the method relies
    on the scientific target properties it is unsuitable for any general
    implementation in the pipeline software for VIMOS. }

   {}


   \maketitle
%

%
%
%
\section{Introduction}
\label{sec:intro}
Integral field units (IFU), combining spectrographic and imaging
capabilities, are used to obtain spatially resolved spectra over a
typically contiguous field-of-view (FoV). These instruments, which offer
an enormous gain in observing efficiency compared to classical
long-slit spectrographs, are now used at all major observatories (e.g.
GMOS at Gemini, SAURON at WHT, and OSIRIS at Keck).  Their often complex
instrument designs and the large amount of information on the detectors
present challenges for the data reduction.  At the VLT, the Visible
Multi-Object Spectrograph (VIMOS) has an IFU containing 6400
microlenses coupled to fibres covering the wavelength range 4000 --
10150\,\AA\/ with a set of six grisms \citep{lefevre}. With the medium
and high-resolution grisms (R=580-2500), the spectra of a single
pseudo-slit will cover the entire length of the CCD, hence only
one pseudo-slit can be used. The IFU head is then partially
masked by a shutter, so that only a square of 40 x 40 fibres is used
yielding either a FoV of $27\arcsec \times 27\arcsec$ at $0\farcs67$
per fibre, or a FoV of $13\arcsec \times 13\arcsec$ at $0\farcs33$ per
fibre. The light is fed to four independent spectrographs each
covering a quarter of the FoV and recording spectra on its own
CCD. The four identical channels of VIMOS are referred to as
quadrants. The VIMOS data reduction pipeline \citep{Izzo,Zanichelli}
removes the instrumental signature from the spectra of each quadrant
and combines data into a single data cube covering the full FoV.

\begin{figure}
\centering
\includegraphics[width=8.5cm]{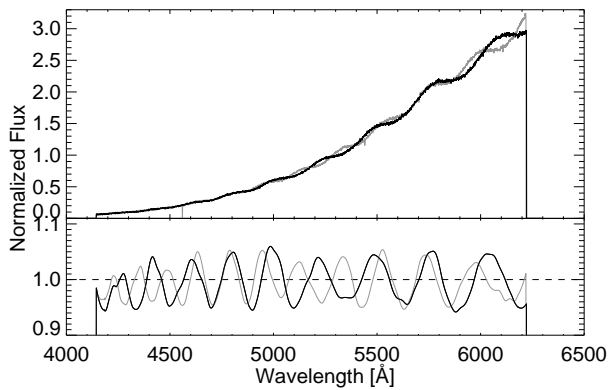}
\caption{The fringe-like pattern in the VIMOS-IFU used with the
  ``HR-Blue'' grism.  Top panel: Flat-field spectrum from a single
  spatial element (fibre) clearly showing the effects of the
  fringe-like pattern (black). Bottom panel: the corresponding,
  normalized correction function for this fibre (black). In gray we
  show the flat-field spectrum and correction function from the same
  spatial element but observed in a different night. }
\label{fig:fringes}
\end{figure}

It has been known for several years \citep[e.g.][VLT VIMOS
manual\footnote{http://www.eso.org/sci/facilities/paranal/instruments/vimos/doc;
  issue 88.0, Sect. 2.8.1}]{jullo} that spectra acquired by the VIMOS-IFU in
conjunction with the ``HR-Blue'' and ``HR-Orange'' grisms exhibit
spectral features that are visually similar to fringes. Fringes normally 
arise from the interference in the CCD detection layer, between
incident light and the light reflected from the interfaces of the CCD
layers. Given the typical thickness of this layer, fringes are
observed at red wavelengths ($>$7000\,\AA). Yet, in the case of the
VIMOS-IFU, features (see Fig.~\ref{fig:fringes}) resembling
fringes are present over the whole wavelength range, suggesting that
they are caused by a different mechanism. For the rest of this
research note we refer to these features as ``fringe-like patterns''.

\citet{jullo} investigated the possible origin of
the fringe-like pattern, concluding that it originates from either 
within the mask at the pseudo-slit level or the imperfect
position of the masks themselves. However, since the
fringe-like pattern varies between exposures they proposed that the
effect probably arises from the flexure of the prisms in the
pseudo-slit. The most likely cause of the fringe-like pattern in the
spectra is the presence of a ``pseudo etalon'', of approximately 5-10
micron thickness, associated with the fibre output prism (H. Dekker,
private communication).

In this research note, we present an empirical method for removing the
fringe-like pattern, which is implemented as an additional step in the
data reduction before stacking individual exposures.  We first personally noticed
the fringe-like pattern during the analysis of our own VIMOS-IFU
observations of nearby early-type galaxies (ETGs) utilizing the
``HR-Blue'' grism (wavelength range 4150-6200\,\AA) with a $27\arcsec
\times 27\arcsec$ FoV.

For this instrument mode and these specific targets, we 
developed and tested the method to remove the fringe-like
pattern. Our targets are typical of nearby ETGs hence have very smooth
surface brightness profiles and cover the entire FoV of the
VIMOS-IFU. The full scientific investigation of these ETGs will be
presented in a future paper (Lagerholm et al. in prep.). Here we only
show our results for the galaxy NGC\,3923, which are used to illustrate our
correction method.

The paper is organized as follows. Sec. \ref{sec:challenges} describes
the challenges arising from the data reduction process. In
Sec. \ref{sec:corrmethod}, our method is used to correct the intensity
differences and the fringe-like pattern is described. In
Sec. \ref{sec:results}, we describe the impact of the corrections on
the scientific analysis and provide a map of the typical strength of
the fringe-like pattern for the VIMOS-IFU. Finally, in
Sec. \ref{sec:discussion} we present our conclusions.


%
%
%
\section{Challenges of the data reduction}
\label{sec:challenges}
The data reduction was mainly performed with the ESO pipeline
(version~2.3.3) using the standard settings described in the pipeline
manual\footnote{http://www.eso.org/sci/software/pipelines/}. The
pipeline processing steps included the subtraction of the median-combined bias image, creation of the spectral extraction mask from a
flat-field image taken immediately following the science exposure, and
wavelength calibration constructed from the HeArNe lamp exposures.
The flux calibration, which is normally performed within the pipeline,
did not deliver satisfactory results for some standard stars.
We therefore, opted to perform the flux calibration outside the
pipeline with $\tt{IRAF}$ using the $\tt{standard, sensfunc}$, and
$\tt{calibrate}$ tasks. This also enables us to combine
several standard stars for the flux calibration, which is impossible
within the ESO pipeline. The standard stars used for the flux
calibration are extracted using the pipeline, where the spectrum is
derived by adding several spatial elements thus diminishing the
signatures from the fringe-like pattern. The final flux calibration
curve, which is derived separately for each quadrant, was constructed
by fitting a third-order polynomial to the standard star data. The
low-order polynomial fit ensures that any signatures of the
fringe-like pattern are removed.

In the following, we describe three data reduction issues that remain
after the standard pipeline procedure: (1) intensity differences
between quadrants, (2) intensity stripes across the complete FoV, and
(3) the fringe-like pattern in the spectral domain.  All three
effects, if not accounted for, can significantly affect the scientific
analysis, especially when the continuum shape and/or the continuum
level play a role.

\subsection{Quadrant-to-quadrant throughput variations}
\label{sec:quadrant}
\begin{figure}
\centering
\includegraphics[width=3.7cm]{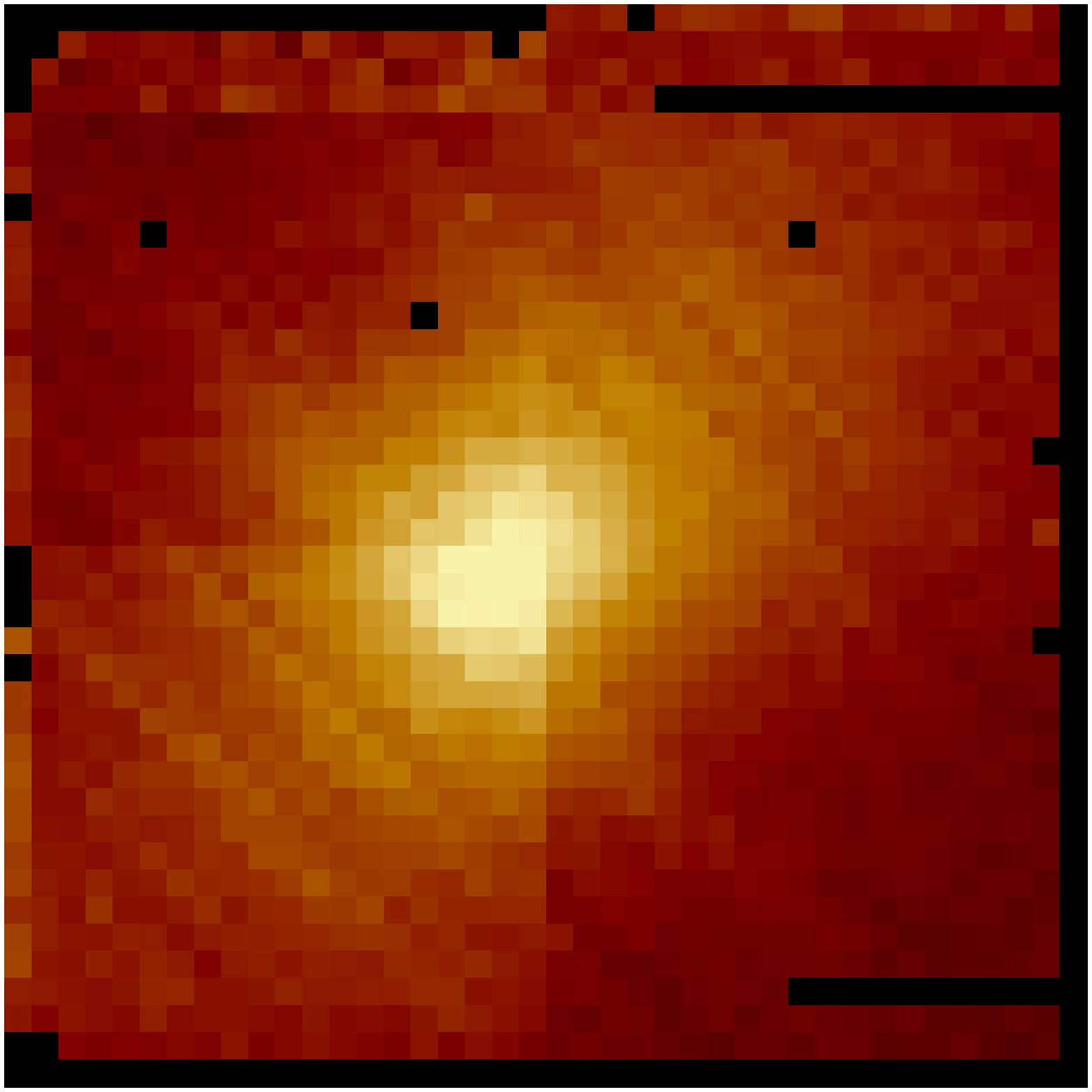}
\includegraphics[width=3.7cm]{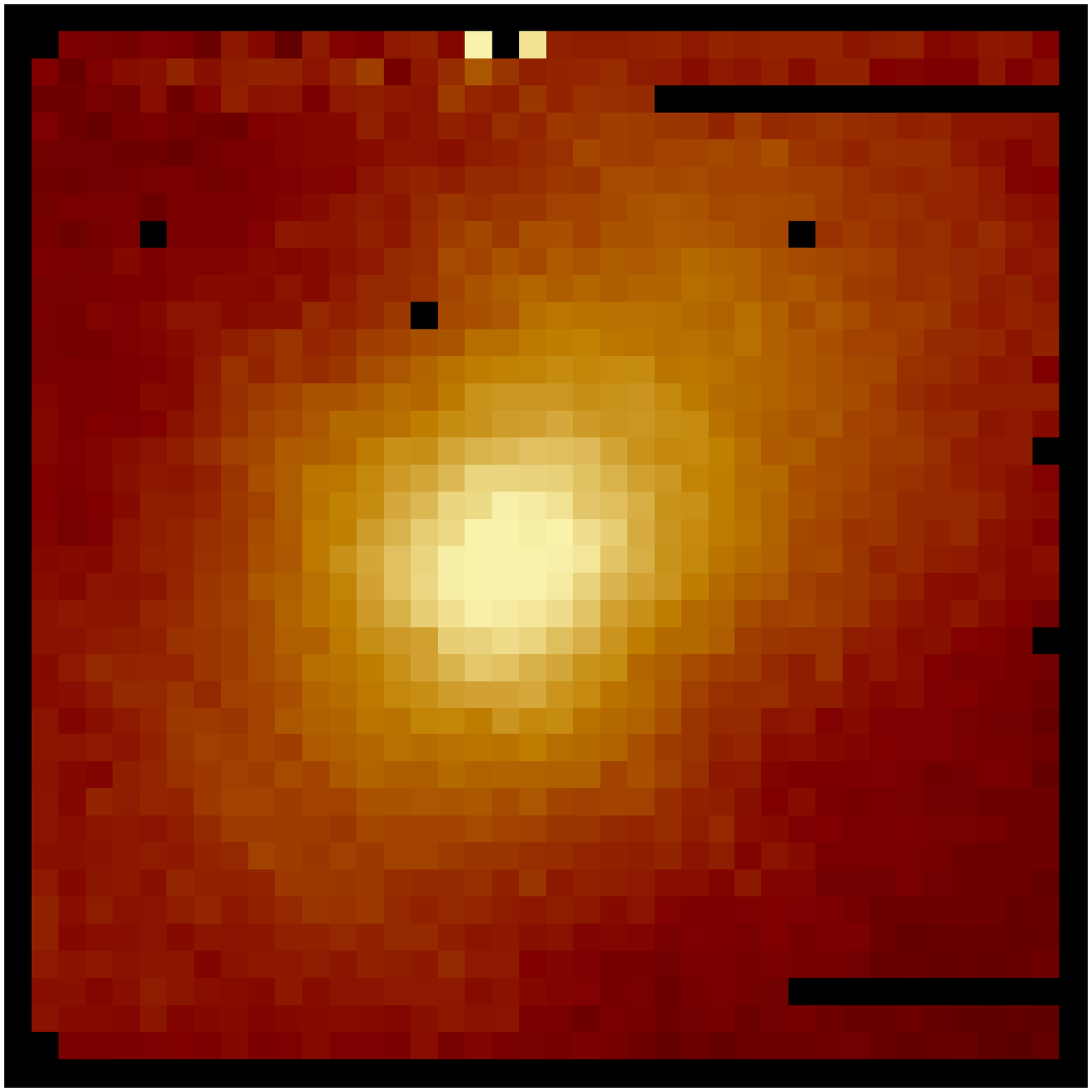}
\includegraphics[width=3.7cm]{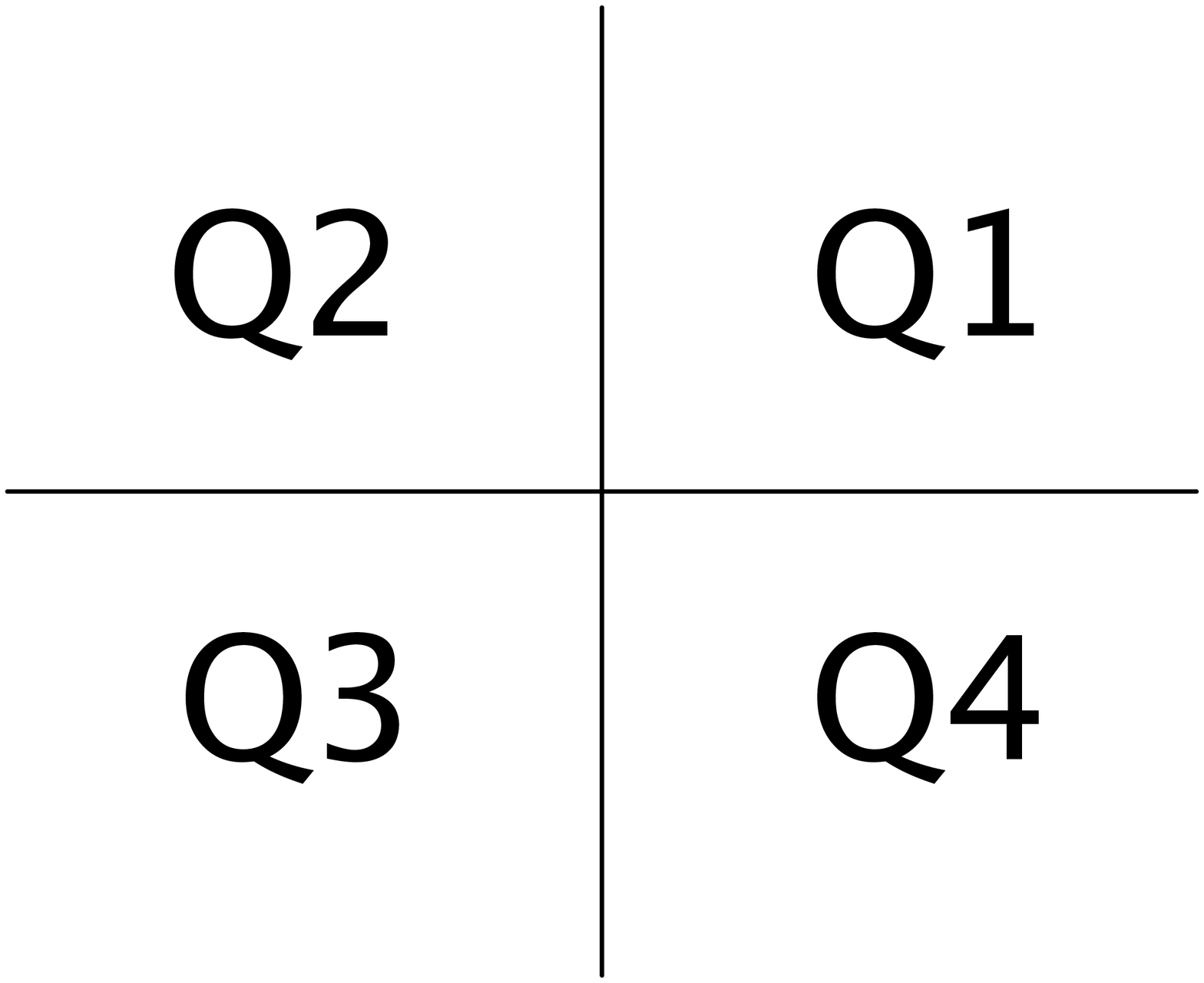}
\includegraphics[width=3.7cm]{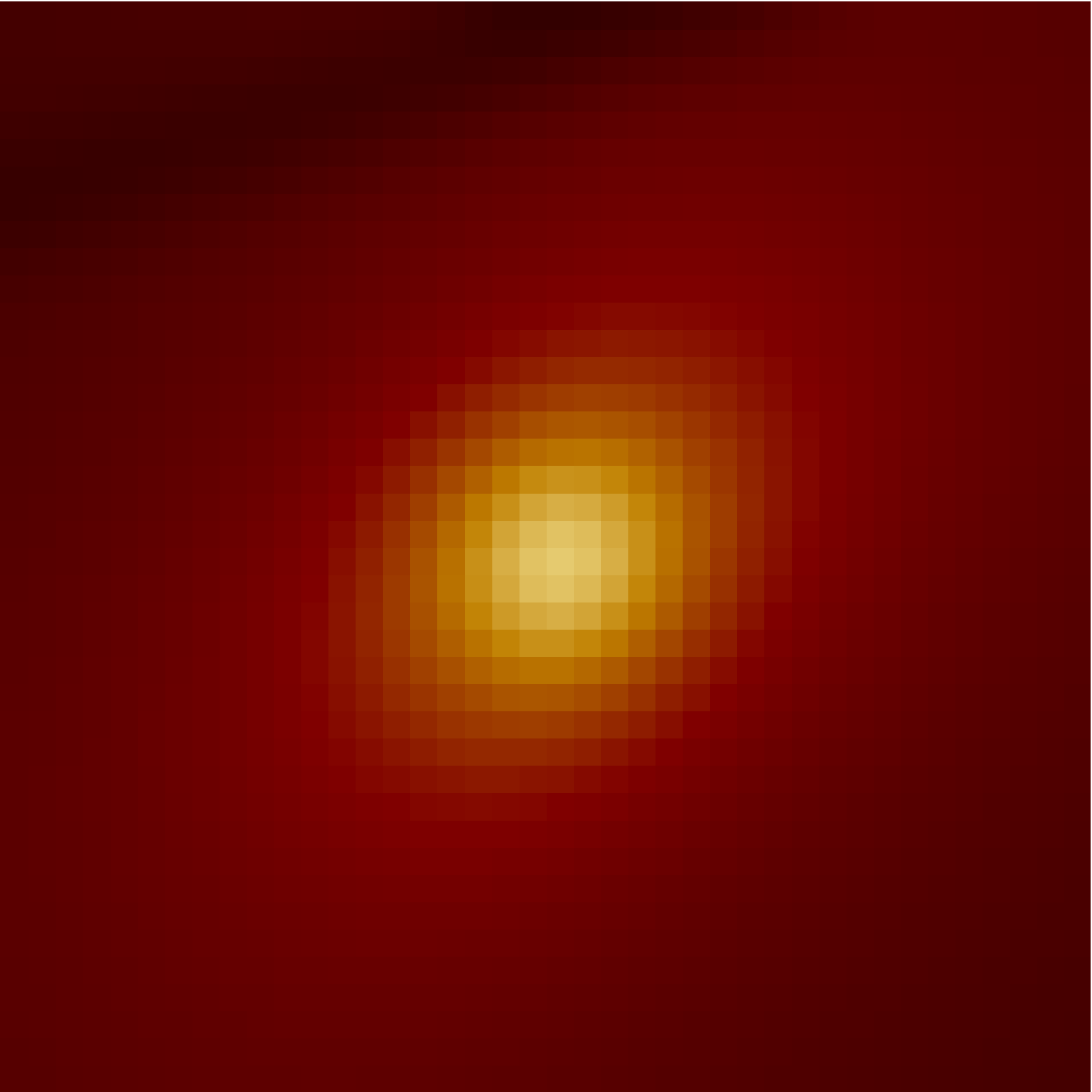}
\caption{Top panels: Reconstructed images derived from a VIMOS-IFU
  data cube showing the central $26\farcs8 \times 26\farcs8$ of the
  early-type galaxy NGC\,3923. Top left: an example of the image
  derived from the original data cube as produced by the pipeline; Top
  right: image derived from the corrected cube. Bottom left: Numbering
  scheme of the four VIMOS-IFU quadrants; Bottom right: For comparison,
  we show an ACS/WFC F606W image of NGC\,3923, binned up to the
  spatial sampling, field-of-view and broadened to $\sim1\farcs2$
  seeing of the VIMOS data-cube. For the three images of NGC\,3923, the
  orientation is north to the right and east to the top.}
\label{fig:cubes}
\end{figure}

Owing to the four independent spectrographs used in VIMOS
\citep{lefevre}, each re-constructed data-cube is composed of four
quadrants (Q1 to Q4; see also Fig. \ref{fig:cubes}). After a standard
pipeline reduction, the quadrants typically have different intensity
levels, as shown in the top left panel of Fig. \ref{fig:cubes}. These
differences (of up to 40\% of the peak intensity for our data) give rise
to sharp drops/increases in intensity along the connecting quadrant
edges. Although these throughput differences between quadrants do not
affect the extraction of e.g. kinematics and line-strength
measurements for individual spectra, a correction is needed when either a
reconstructed image is analyzed or regions across the quadrant
edges are binned to reach a specific signal-to-noise ratio (S/N) requirement.

\subsection{Intensity stripes}
\label{sec:intstripes}
In addition to the quadrant-to-quadrant intensity variations, the
reconstructed images from the VIMOS-IFU cubes have diagonal intensity
stripes across the FoV (Fig.~\ref{fig:cubes}; top left panel). The
intensity variations in these diagonal stripes reach up to $\sim$15\%
in our data. When comparing individual spectra located at the peak of
a given stripe with those next to it, the spectra associated with
the stripe have an overall higher flux level over the whole wavelength
range.  If these stripes were caused by the instrument, one
would expect to find discontinuities at the quadrant borders, since the four
quadrants originate from independent spectrographs. That the
stripes seem to propagate all the way through the quadrant borders implies that they are caused by a pipeline reduction problem, the underlying cause of which is
presently unknown.

\subsection{Fringe-like patterns}
\label{sec:fringe}
\begin{figure}
\centering
\includegraphics[width=8cm]{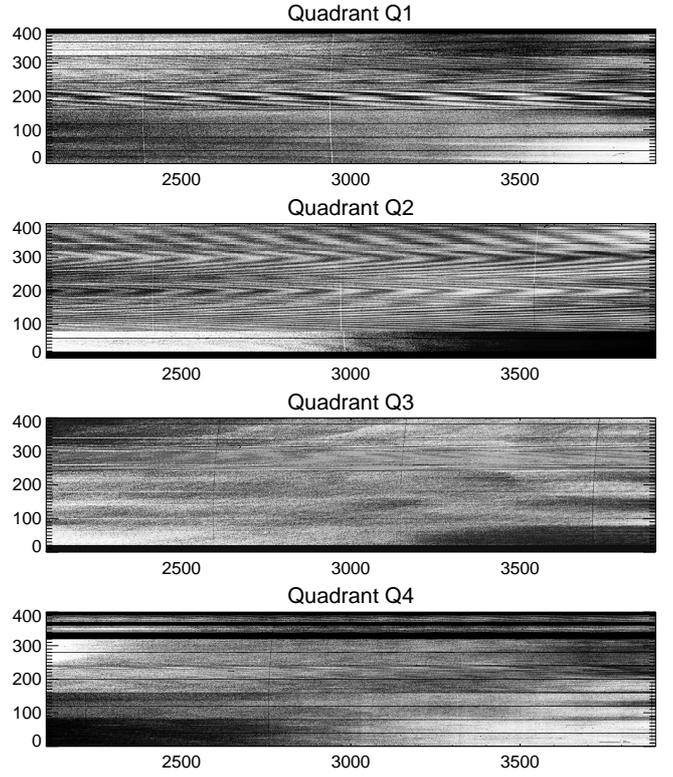}
\caption{Variation in the fringe-like pattern for extracted flat-field
  spectra where the median lamp spectrum has been removed. Each row
  represents one fibre and wavelength increases from left to right. A
  subset of the wavelength range, covering pixels 2100 to 3900 and
  thus corresponding to about 5130\,\AA, to 5675\,\AA, is shown. The example
  spectrum shown in Fig.~\ref{fig:fringes} is taken from quadrant Q3,
  where only one fibre module shows a significant fringe-like
  pattern.}
\label{fig:rawflat}
\end{figure}

\begin{figure}
\centering
\includegraphics[width=8.5cm]{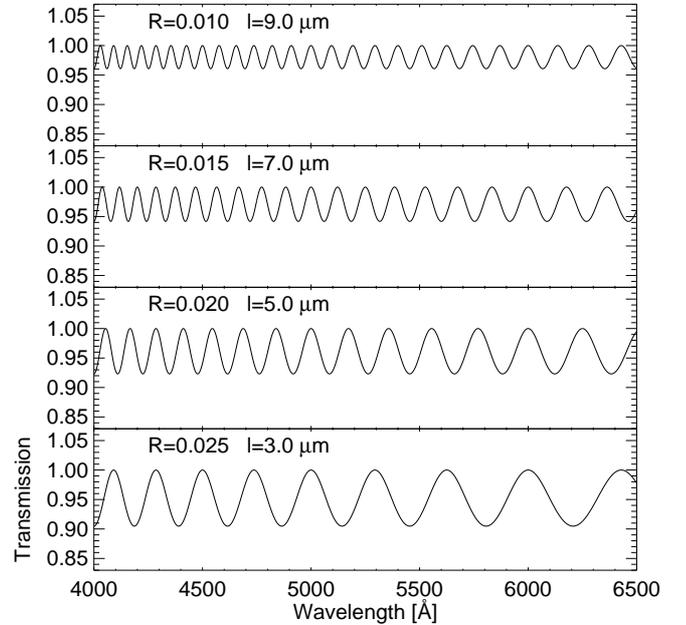}
\caption{The transmission of an etalon as a function of wavelength for
  a few values of reflectance $R$ and etalon thickness $l$. The values
  are indicated inside each panel.}
\label{fig:etalon}
\end{figure}

More than half of the spectra in a VIMOS data-cube are significantly
affected by the fringe-like pattern (see Fig.~\ref{fig:fringes}). The
amplitude and the frequency of the pattern is not randomly distributed
between fibres but shows a clear connection to individual fibre
modules. We note that the fringe-like pattern is also present in the raw data, thus is not an artifact created by the extraction process of the spectra. The fringe pattern shifts monotonically inside a given module
but strong breaks appear between fibre modules as illustrated in
Fig.~\ref{fig:rawflat}. In quadrant Q2, almost all spectra are affected,
while in the other quadrants fibre modules have a negligible or
no fringe-like pattern at all. For each fibre, the fringe-like pattern
resembles to first order the transmittance function of an etalon with
a reflectance of order $1-3$\% and a thickness of order $3-10\,\mu$m
(see Fig.~\ref{fig:etalon} for examples).  However, the variation in
the amplitude with wavelength and the deviations from a simple etalon
transmission function suggest that more than one layer and complex
reflectance variations with wavelength may be present in reality.
  
We note that the fringe pattern varies considerably
with time (see Fig.~\ref{fig:fringes}; gray lines).  Even flat fields
taken during the night and adjacent to the science exposure typically
do not show the same fringe-like pattern as in the science data. This
can be explained by the change in the instrument rotator angle and
associated flexure in the instrument. During the science exposure, the
rotator angle of the instrument changes in order to follow the target,
leading to a different bending of the fibres and therefore a variation in
the flexure of the prisms in the pseudo-slit and the fringe-like
pattern.  The flat-field taken at the end of the science exposure only
reflects the pattern at the final rotator angle. Owing to hysteresis
effects, returning to the same rotator angle also leads to a different
fringe-like pattern \citep{jullo}.

We further note that the observed fringe-like pattern has a frequency
and amplitude similar to stellar absorption features in our target
galaxy spectra and thus simple filtering with Fourier techniques is unsuitable for removing the effect. For the most affected spectra,
the pattern accounts for $\sim$13\% difference in intensity,
peak-to-valley (hereafter PTV; see also Sec. \ref{sec:FringeMap}).

\section{The correction method}
\label{sec:corrmethod}
After the extraction and wavelength calibration of spectra using the
VIMOS pipeline and flux calibration within IRAF, we correct for the
quadrant-to-quadrant intensity differences. We assume that the
intensity correction is constant with wavelength and uniform within each quadrant, and re-normalize
the quadrants Q1, Q3, and Q4 to quadrant Q2 (see Fig.~\ref{fig:cubes}; top
right panel). The re-normalization is done by comparing the intensity
levels of the neighboring pixels at the quadrant borders and taking a
first-order gradient in intensity across the borders into account.

We can now turn our attention to the fringe-like pattern in the
spectra and the intensity stripes seen in the reconstructed image.
For both, the correction of the fringe-like pattern and the
intensity-stripes, the underlying main assumption is that, to first
order, the effects are localized, i.e. they are independent among
even neighboring spectra. This is a reasonable assumption if the
effects are caused by an imperfect fixation of the fibre to the prism
\citep{jullo} creating a ``pseudo etalon'' associated with the fibre
output prism.

\begin{figure}
\centering
\includegraphics[width=9cm]{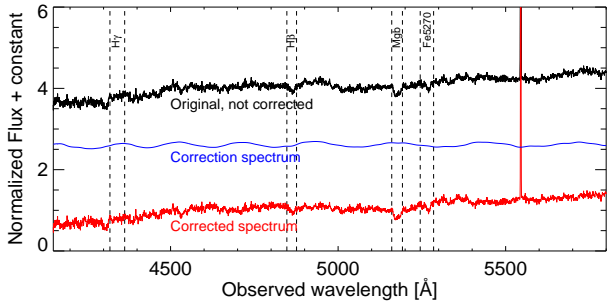}
\caption{Example of the correction made for each individual spectrum.
  Here one of the most affected spectra for the galaxy NGC\,3923 (not
  sky-subtracted) is shown. The top black spectrum represent the
  original spectrum at one spatial position. The middle blue spectrum
  shows the correction spectrum derived with our method. The bottom
  red spectrum shows the corrected spectrum. The black dashed vertical
  lines indicate the central bandpasses of the line-strength indices
  H$\gamma_{A}$, H$\beta$, Mg\,$b$, and Fe5270.}
\label{fig:spectrafringe}
\end{figure}

For each spectrum in the data-cube, we calculate the {\em median}\/
spectrum from the nearest eight spatially surrounding spectra. This median spectrum is,
if all eight spectra have different fringe-like patterns, to first
order unaffected by the pattern. Since the spectral properties of our targets
 vary relatively slowly as a function of
spatial position and the signal in the neighboring spectra is
correlated due to natural seeing effects, this median spectrum can be
used as an approximation of the underlying, ``true'' spectrum in the
central pixel.  The ratio of the median spectrum to that of the
central pixel will provide an estimate of the fringe-like pattern,
i.e. a correction spectrum that by construction has a mean of about one.
Owing to the typically limited S/N of the correction spectrum we decided
to smooth the correction spectrum within $\tt{IDL}$ using the
$\tt{lowess}$ function, which is part of The $\tt{IDL}$ Astronomy
User's Library \citep{IDLAstronomy}.  We applied a smoothing function
using a second-order polynomial for each step of 150 pixels in our
case. The resulting correction spectrum preserves the overall shape
and amplitude of the fringe-like pattern with negligible noise at
smaller wavelength scales (see Fig. \ref{fig:spectrafringe}).

The fringe-like pattern is then removed from the original data-cube by
dividing each spectrum by the corresponding smoothed correction
spectrum. This procedure is done for all the spectra in the cube,
except for the outermost corner spectra and spectra in the
neighborhood of regions with dead fibres (shown as black dots or black
stripes in the reconstructed IFU image in Fig. 2).  Spectra for which
we could not derive a correction spectrum based on the median of at
least five or more surrounding spectra were set to zero.

A drawback of the above-described procedure is the effective smoothing
that is applied over the scale length of roughly three spatial
elements. In particular, at the center of the target galaxies this is
a non-desirable effect that has to be accounted for in any subsequent
analysis. However, the question arises of whether the median of the eight
surrounding spectra provides a good enough estimate of the
underlying spectrum or one would need to combine even more spectra
to obtain a sufficiently high quality correction spectrum.

To test the effect of increasing the number of spatial
elements used to derive the median spectrum, we re-constructed a
data-cube for an internal flat-field exposure. Here it can be safely
assumed that over the scale length of several pixels the underlying,
uncontaminated spectrum does not change. In our test, we doubled the
number of spectra used for the median spectrum, by considering both the eight
surrounding spectra plus eight spectra, equally spaced, from the next
closest ring of spectra in the data cube.

\begin{figure}
\centering
\includegraphics[width=9cm]{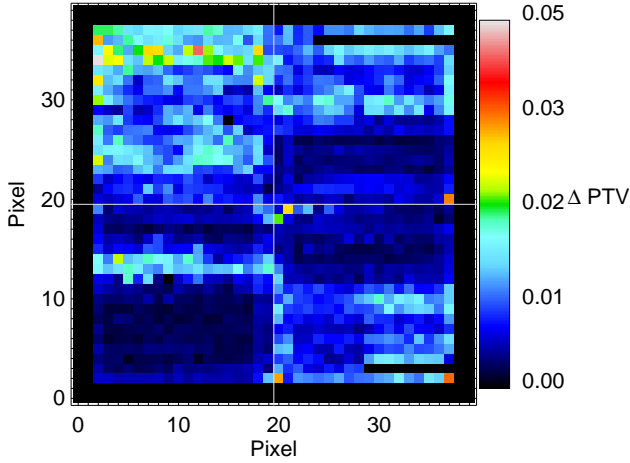}
\caption{The maximum differences between the fringe-like patterns,
  created using both 8 and of the 16 surrounding
  spectra, in the wavelength range $5188-5620$\,\AA. The white solid
  lines indicate the borders of the different quadrants. The quadrants
  are positioned as shown in the bottom left panel of Fig. \ref{fig:cubes}.}
\label{fig:flatfringe}
\end{figure}

Fig.~\ref{fig:flatfringe} shows the maximum difference between the
correction spectrum constructed from 8 and 16 neighboring spectra
for the whole VIMOS-IFU FoV. Except for a few outliers, especially in
the top of quadrant Q2, which show differences up to 5\%, most of the
correction spectra show difference of less than 0.6\%. We conclude
that the median of eight spectra is already sufficient to produce a
correction spectrum with an accuracy of about 0.6\%. Given that our
typical corrections are on the order of 6\% and that we combine
several exposures to form the final data cube for analysis, this
accuracy appears sufficient. Furthermore, a visual inspection of the
differences between taking the median of 8 or 16 spectra indicates
that the central wavelength range, containing the most important
absorption features, is well-behaved, while extreme differences are
often found at the edges of the wavelength range.

%
%
%
\section{Results}
\label{sec:results}
We first describe here the potential consequences of an
imperfect data reduction (Sec.~\ref{sec:corrmethod}) for a typical
kinematical and stellar population analysis of early-type galaxies. We
then demonstrate in Sec.~\ref{sec:application}, by using the example
of our VIMOS-IFU observations of the ETG NGC\,3923, the extent to which  the
proposed correction method removes the effects.  Finally, we provide a
map of the VIMOS-IFU FoV showing the regions that are most affected
by the fringe-like pattern (see Sec.~\ref{sec:FringeMap}).

\subsection{Science impact}
\label{sec:sci_impact}
\begin{figure}
\centering
\includegraphics[width=9cm]{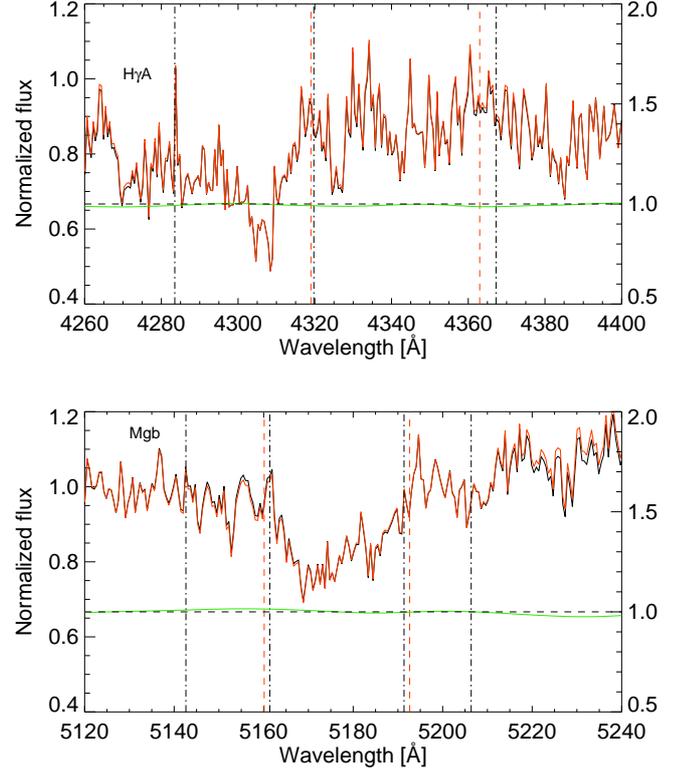}
\caption{Difference between a spectrum corrected for the fringe-like
  pattern (red solid line) and a spectrum that was not corrected
  (black solid line). The green solid line shows the correction
  function across the wavelength range where the scaling is given
  at the right y-axis.  The line strength is measured between the red
  dashed lines and the continuum band-passes between the dashed-dotted
  vertical lines. The line-strength measurements for these two lines
  in the uncorrected and corrected spectrum are for H$\gamma_{A} =
  -4.79 \pm 0.33$\,\AA, and $-5.07 \pm 0.37$\,\AA, and for Mg\,$b =
  5.33 \pm 0.28$\,\AA, and $5.25 \pm 0.32$\,\AA, respectively. }
\label{fig:linespectra}
\end{figure}

Both the intensity differences and the fringe-like pattern, if uncorrected for, will affect the science derived from the data.  When
considering the whole spectral range (see Fig.
\ref{fig:spectrafringe}), it can be hard to see the effects that the
corrections have on the spectra. In Fig. \ref{fig:linespectra}, we 
therefore zoomed into the spectral range of two absorption features,
H$\gamma$ and Mg\,$b$, used for the line-strength analysis of the
stellar populations in NGC\,3923.  In the Lick/IDS system
\citep[][]{tra98}, absorption line-strengths are measured by indices,
where a central feature bandpass is flanked to the blue and red by
pseudo-continuum bandpasses (see Fig.~\ref{fig:linespectra}).  The
mean level of the two pseudo-continuum regions is determined
independently on each side of the feature bandpass and a straight line
is drawn through the midpoint of each one. The difference in flux
between this line and the observed spectrum within the feature
bandpass determines the index. In our example, shown in
Fig.~\ref{fig:linespectra}, it is clearly demonstrated that the
fringe-like pattern correction changes the continuum level of the
spectra on a scale relevant to the determination of the
line-strength. In particular, the pseudo-continuum level used in the
line-strength analysis, will be located wrongly. For the example
spectrum shown in Fig. 5, the Mg\,$b$\/ and H$\gamma_{A}$ indices
change by $-0.08$\,\AA, and $-0.28$\,\AA, respectively (uncorrected $-$
corrected spectrum).

The example shown in Fig. \ref{fig:linespectra} represents only one
realization of the fringe-like pattern and it can be easily seen how
different patterns can lead to a whole range of changes in the
line-strength measurements (positive and negative
changes). Furthermore, velocity and velocity dispersion measurements
can also be affected,although to a smaller extent since the measurements are typically obtained
from a wavelength range covering several wiggles/periods of the
fringe-like pattern.

\begin{figure}
\centering
\includegraphics[width=9cm]{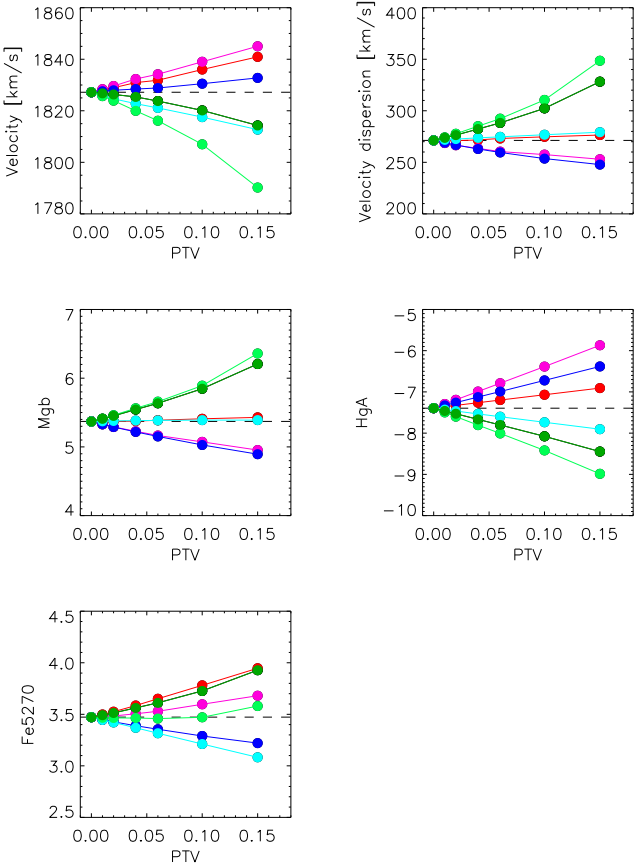}
\caption{Changes in velocity, velocity dispersion and the
  line-strengths, Mg\,$b$\/, H$\gamma_{A}$ and Fe5270 depending on the
  PTV of the fringe-like pattern (sinusoidal). Different
  colors indicates different phase shifts of the sinusoidal, equally
  separated between 0 and 2$\pi$. The shift in color goes from dark green,
  red, magenta, blue, light blue, light green, back to dark green. The
  black dashed lines indicates the correct values given by the fringe
  free galaxy spectrum.}
\label{fig:genfringe}
\end{figure}

To evaluate the effects more quantitatively, we set up simple
simulations.  The fringe-like pattern is not a simple sinusoidal
function, but we approximate it with a sinusoidal with the mean
frequency of the fringe-like pattern and vary the phase and amplitude
covering the full range seen in the real data. This sinusoidal is
multiplied by a fringe-free galaxy spectrum (a galaxy spectrum already
corrected for the fringe-like pattern) to simulate the effects of the
fringe-like pattern. In our simulations, we changed the amplitude (PTV
between 0\% and 15\%) and phase (between 0 and 2$\pi$) of the sinusoidal
each in seven steps and measured for each spectrum the velocity,
velocity dispersion, and line-strengths. The results are summarized in
Fig.~\ref{fig:genfringe}. As expected, the exact location of the
fringe-like pattern determines whether a given quantity is changed in a
positive or negative direction. Even for large amplitudes, a
negligible change is possible when the effects of the fringe-like
pattern cancel out. However, the exact phase for this configuration
depends on the quantity under consideration.

For a typical amplitude of 5\% (PTV = 0.10), the velocities can be
affected by up to $\pm 20$\,\kms, the velocity dispersion up to $\pm
40$\,\kms\/, the Mg\,$b$\/ index by up to $\pm 0.5$\,\AA, the
H$\gamma_{A}$ index up to $\pm 1$\,\AA\,, and the Fe5270 index up to
$\pm 0.3$\,\AA.

Overall, we conclude that for these typical correction-spectrum
amplitudes (see also Sec.~\ref{sec:FringeMap}), the fringe-like
pattern can have a significant influence on the scientific analysis,
especially for kinematical and line-strength measurements. We
emphasize that the above numbers are only valid for an individual
exposure and that the combination of several exposures or fibres will
significantly mitigate the problem.

\subsection{Application to our data}
\label{sec:application}
Overall, our correction method provides significant improvements
compared to the results obtained from uncorrected data. In the
reconstructed image, the intensity differences, both between quadrants
and between individual spectra seen as stripes, are now considerably
smaller (on average $<$5\%; see Fig. \ref{fig:cubes}; top left
panel). The overall morphology of the science target (e.g. the
isophotes) is now in much closer agreement with a direct image,
although it does not reach its level of quality (see
Fig.~\ref{fig:cubes}; bottom right panel).

\begin{figure}
\centering
\includegraphics[width=9cm]{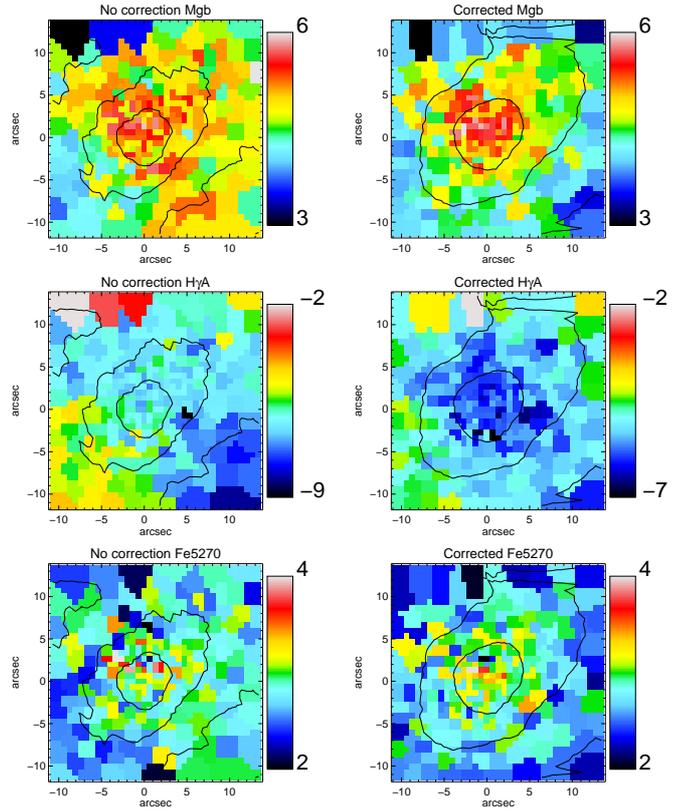}
\caption{Line-strength maps for NGC\,3923 over Mg\,$b$, H$\gamma_{A}$,
  and Fe5270. In the three plots to the left, the cubes are not
  corrected and in the three plots to the right the cubes are
  corrected for the fringe-like patterns. The spectra used for the
  line-strength analysis are binned, using Voronoi binning
  \citep{CapCopin}, in order to have at least a S/N of 60. The
  velocities and velocity dispersions are calculated using pPFX
  \citep{CappellariEmsellem} and are used in the line-strength
  measurements. The line strengths are measured using the same
  principle as in \citet{Kuntschner2006}.}
\label{fig:linemap}
\end{figure}

However, the most important part of the correction is the removal of
the fringe-like pattern from the spectra. In Fig.~\ref{fig:linemap}, we
show the line-strength maps for the Mg\,$b$, H$\gamma_{A}$, and Fe5270
indices of NGC\,3923, for both the fringe-like pattern corrected
and uncorrected cubes. Each cube is the combination of five
individual exposures. In the case of the corrected cube, each
individual exposure was corrected before combination.

The overall improvement in quality can be directly seen by comparing
the uncorrected and corrected line-strength maps in
Fig.~\ref{fig:linemap}. The maps constructed with the corrected cubes
have metal index values (Mg\,$b$\/ and Fe5270) that peak towards the
center, while the maps for the uncorrected cubes show a more
scattered morphology and are not as smooth as expected for an ETG
\citep[e.g.][]{Kuntschner2006}. However, the expected differences
between the corrected and non-corrected line-strength maps are not as
large as one would expect from our simulations described in
Sec.~\ref{sec:sci_impact}. This can be understood by considering
that the final cube was compiled from the combination of five exposures,
which helped to average out some of the fringe-like pattern.

The effects of the fringe-like pattern scale with the signal in the
data since it is a multiplicative effect. Therefore, a low
signal-to-noise exposure, e.g. a sky exposure, will not display a
prominent fringe-like pattern. The fringe-like pattern will be hidden
in the noise, and we are unable to correct them using our proposed
method. However, for the sky subtraction of the galaxies we use a
median over the whole sky cube, thus the sky spectrum will essentially
be free from the fringe-like pattern.

For science targets for which our proposed method is unsuitable
(e.g. low signal-to-noise or highly variable spectral properties
across the FoV), one can reduce the fringe-like pattern by averaging
several, slightly dithered, exposures. The appropriate number of
exposures depends on the science goals and the target itself,
but as a general guideline we would recommend obtaining on the order of eight
exposures thereby mimicking the averaging effect of our proposed
method.

\subsection{Spatial map of the fringe-like pattern effect}
\label{sec:FringeMap}
\begin{figure}
\centering
\includegraphics[width=9cm]{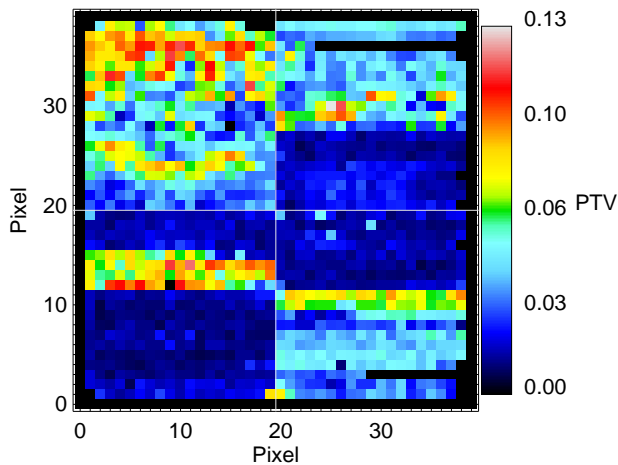}
\caption{Spatial map of the maximum variations (PTV) in the correction
  spectra as derived from a flat-field and the wavelength range
  $5188-5620$\,\AA.  Regions with a strong fringe-like pattern are
  clearly visible and related to certain fibre models (see also
  Fig.~\ref{fig:rawflat}).  The white solid lines indicate the borders
  of the quadrants. The quadrants are positioned as shown in
  Fig. \ref{fig:cubes}, bottom left panel.}
\label{fig:PTVmap}
\end{figure}

The four quadrants are not all equally affected by the fringe-like
pattern -- there are variations both between and within
the quadrants. Fig.~\ref{fig:PTVmap} shows the maximum PTV values per
spatial element, derived from the correction spectra of a flat field.
Regions with a prominent fringe-like pattern, connected to individual
fibre modules (see also Fig.~\ref{fig:rawflat}) are clearly visible.
While the precise phase and amplitude of the fringe-like pattern
varies between exposures, the spatial regions that are significantly
affected in a VIMOS-IFU cube can be clearly characterized.  For example,
quadrant Q2 is the most affected quadrant. The other three quadrants
feature regions that are much less or not at all affected.  The
maximum PTV values in a cube reach 0.13, while the mean value for the
affected fibre models is about 0.06, i.e. an amplitude of 3\%.

%
%
\section{Conclusions}
\label{sec:discussion}
We have shown that additional data reduction
steps, to complement the standard ESO data reduction pipeline recipes, can
improve the quality of the reduced spectra and therefore the science
analysis of VIMOS-IFU data obtained with the ``HR-Blue'' grism. The
correction of quadrant-to-quadrant intensity variations and intensity
stripes leads to significantly improved reconstructed images. More
than half of the VIMOS-IFU spectra obtained with the ``HR-Blue'' grism
show, across the full wavelength range, spectral features visually
similar to fringes.  Unfortunately, the fringe-like pattern is unstable with time and cannot therefore be removed in the flat-fielding
process.  Our proposed empirical method for minimizing the effects of
the fringe-like pattern enables a meaningful kinematical and
line-strength analysis for nearby early-type galaxies as demonstrated
for the case of NGC\,3923.  The combination of several exposures into
a combined cube also mitigates the fringe-like pattern, although, our
newly proposed method working on individual exposures before
combination, provides better results.  We note, that the proposed
correction method is only tested for and will work with objects
similar to our science targets (nearby early-type galaxies) that
cover the whole VIMOS-IFU FoV and display slowly varying spectral
properties over the FoV. When dealing with observations that have
strongly varying background or low intensities a different
approach should preferably  be adopted. In these cases it may be better to rely on
the combination of several individual exposures to reduce the effects
of the fringe-like pattern or to preform a full modeling of the
fringe-like pattern.


\begin{acknowledgements}
  We would like to thank H. Dekker for useful comments on the possible
  origin of the fringe-like pattern. We thank the anonymous referee for useful comments. MC acknowledges support from a
  Royal Society University Research Fellowship. RMcD is supported by
  the Gemini Observatory, which is operated by the Association of
  Universities for Research in Astronomy, Inc., on behalf of the
  international Gemini partnership of Argentina, Australia, Brazil,
  Canada, Chile, the United Kingdom, and the United States of America.
\end{acknowledgements}

\bibliographystyle{aa} 
\bibliography{references} 

\end{document}